\font\bbb=msbm10                                    

\overfullrule=0pt

\def\C{\hbox{\bbb C}}  
  
\def\R{\hbox{\bbb R}}  
\def\Z{\hbox{\bbb Z}}

\def\FoCM{{\sl Found.\ Computational Math.}}
\def\FortP{{\sl Fortsch.\ Phys.}}

\def\JMO{{\sl J. Mod.\ Optics\/}}

\def\JPA{{\sl J. Phys.\ A:  Math.\ Gen.}}

\def\Naw{{\sl Naturwissenschaften}}

\def\PR{{\sl Phys.\ Rev.}}
\def\PRA{{\sl Phys.\ Rev.\ A\/}}

\def\PRL{{\sl Phys.\ Rev.\ Lett.}}
\def\PRSLA{{\sl Proc.\ Roy.\ Soc.\ Lond.\ A\/}}

\def\PhT{{\sl Phys.\ Today\/}}

\def\ZP{{\sl Z. Physik\/}}

\def\dajm{\hbox{D. A. Meyer}}

\def\mhfreedman{\hbox{M. H. Freedman}}
\def\nrwallach{\hbox{N. R. Wallach}}

\def\md{\hbox{\mhfreedman\ and \dajm}}

\def\dn{\hbox{\dajm\ and \nrwallach}}

\def\gottesman{\hbox{D. Gottesman}}

\def\schrodinger{\hbox{E. Schr\"odinger}}

\def\vidal{\hbox{G. Vidal}}

\catcode`@=11
\newskip\ttglue

   \font\ninerm=cmr9    \font\eightrm=cmr8   \font\sixrm=cmr6
  \font\ninebf=cmbx9   \font\eightbf=cmbx8  \font\sixbf=cmbx6
  \font\nineit=cmti9   \font\eightit=cmti8  
  \font\ninesl=cmsl9   \font\eightsl=cmsl8  
  \font\ninemi=cmmi9   \font\eightmi=cmmi8  \font\sixmi=cmmi6

\font\bigten=cmr10 scaled\magstep2 

\def\ninepoint{\def\rm{\fam0\ninerm}%
  \textfont0=\ninerm \scriptfont0=\sixrm
  \textfont1=\ninemi \scriptfont1=\sixmi
  \textfont\itfam=\nineit  \def\it{\fam\itfam\nineit}%
  \textfont\slfam=\ninesl  \def\sl{\fam\slfam\ninesl}%
  \textfont\bffam=\ninebf  \scriptfont\bffam=\sixbf
    \def\bf{\fam\bffam\ninebf}%
  \tt \ttglue=.5em plus.25em minus.15em
  \normalbaselineskip=11pt
  \setbox\strutbox=\hbox{\vrule height8pt depth3pt width0pt}%
  \normalbaselines\rm}

\def\eightpoint{\def\rm{\fam0\eightrm}%
  \textfont0=\eightrm \scriptfont0=\sixrm
  \textfont1=\eightmi \scriptfont1=\sixmi
  \textfont\itfam=\eightit  \def\it{\fam\itfam\eightit}%
  \textfont\slfam=\eightsl  \def\sl{\fam\slfam\eightsl}%
  \textfont\bffam=\eightbf  \scriptfont\bffam=\sixbf
    \def\bf{\fam\bffam\eightbf}%
  \tt \ttglue=.5em plus.25em minus.15em
  \normalbaselineskip=9pt
  \setbox\strutbox=\hbox{\vrule height7pt depth2pt width0pt}%
  \normalbaselines\rm}

\def\sfootnote#1{\edef\@sf{\spacefactor\the\spacefactor}#1\@sf
      \insert\footins\bgroup\eightpoint
      \interlinepenalty100 \let\par=\endgraf
        \leftskip=0pt \rightskip=0pt
        \splittopskip=10pt plus 1pt minus 1pt \floatingpenalty=20000
        \parskip=0pt\smallskip\item{#1}\bgroup\strut\aftergroup\@foot\let\next}
\skip\footins=12pt plus 2pt minus 2pt
\dimen\footins=30pc

\def\ie{{\it i.e.}}

\def\etc{{\it etc.}}

\def\Proposition{P{\eightpoint ROPOSITION}}

\def\endproof{\vrule height6pt width4pt depth2pt}

\def\and{{\eightpoint AND}}

\def\Htwon{(\C^2)^{\otimes n}}
\def\Hless{(\C^2)^{\otimes n-1}}

\def\SchrodingerA{1}
\def\EPR{2}
\def\BBPS{3}
\def\Vmono{4}
\def\Vpure{5}
\def\Bohm{6}
\def\GHZ{7}
\def\Mermin{8}
\def\LindenPopescu{9}
\def\GRB{10}
\def\CarteretSudbery{11}
\def\Sudbery{12}
\def\AACJLT{13}
\def\AAJT{14}
\def\threequbits{15}
\def\CKW{16}
\def\mono{17}
\def\Dirac{18}
\def\Heisenberg{19}
\def\Bethe{20}
\def\OConnorWootters{21}
\def\Gottesman{22}
\def\BDSW{23}
\def\LMPZ{24}
\def\CGL{25}
\def\Shor{26}
\def\rpp{27}
\def\BarnumLinden{28}
\def\Grover{29}
\def\WongChristensen{30}
\def\EisertBriegel{31}
\def\VidalWerner{32}
\def\Preskill{33}
\def\BriegelRaussendorf{34}
\def\ABV{35}

\magnification=1200

\input epsf.tex

\dimen0=\hsize \divide\dimen0 by 13 \dimendef\chasm=0
\dimen1=\chasm \multiply\dimen1 by  6 \dimendef\halfwidth=1
\dimen2=\chasm \multiply\dimen2 by  7 \dimendef\secondstart=2


\line{\hfill                                              1 June 2001}
\line{\hfill                                         quant-ph/0108104}
\vfill
\centerline{\bf\bigten GLOBAL ENTANGLEMENT}
\medskip
\centerline{\bf\bigten IN MULTIPARTICLE SYSTEMS}
\bigskip\bigskip
\centerline{\bf David A. Meyer and Nolan R. Wallach}
\bigskip
\centerline{\sl Project in Geometry and Physics}
\centerline{\sl Department of Mathematics}
\centerline{\sl University of California/San Diego}
\centerline{\sl La Jolla, CA 92093-0112}
\centerline{$\{${\tt dmeyer}, {\tt nwallach}$\}${\tt @math.ucsd.edu}}
\vfill
\centerline{ABSTRACT}
\bigskip
\noindent We define a polynomial measure of multiparticle entanglement
which is scalable, \ie, which applies to any number of 
spin-${1\over2}$ particles.  By evaluating it for three particle 
states, for eigenstates of the one dimensional Heisenberg 
antiferromagnet and on quantum error correcting code subspaces, we 
illustrate the extent to which it quantifies global entanglement.  We
also apply it to track the evolution of entanglement during a quantum
computation.

\bigskip\bigskip
\noindent 2001 Physics and Astronomy Classification Scheme:
                   03.65.Ud, 
                   05.30.-d, 
                   03.67.Lx. 

\noindent 2000 American Mathematical Society Subject Classification:
                   81P68,    
                   81R05.    
\smallskip
\global\setbox1=\hbox{Key Words:\enspace}
\parindent=\wd1
\item{Key Words:}  multiparticle entanglement, lattice spin models, 
                   quantum algorithms.

\vfill
\eject

\headline{\ninepoint\it Global entanglement   \hfill Meyer \& Wallach}

\parskip=10pt
\parindent=20pt

Although entanglement has been recognized as a remarkable feature of
quantum mechanics since Schr\"odinger introduced the word
[\SchrodingerA] in response to Einstein, Podolsky and Rosen's famous
paper [\EPR], it remains only incompletely understood.  In fact, for
more than two particles---even of only spin-${1\over2}$---there is no
complete classification of entanglement.  To be more precise, a 
{\sl measure of entanglement\/} is a function on the space of states 
of a multiparticle system which is invariant under local unitary 
operators, \ie, unitary transformations on individual particles.  
Thus a complete classification of entanglement for a multiparticle 
system is a characterization of all such functions.  Under the most 
general local operations assisted by classical communication (LOCC 
[\BBPS]), entanglement can change.  A measure of entanglement which 
decreases under LOCC is called an {\sl entanglement monotone\/} 
[\Vmono].

On two particle pure states, for example, all measures of entanglement
are functions of the eigenvalues of the reduced density matrix 
(obtained by tracing the density matrix for the whole system over the 
degrees of freedom of one of the particles), and sums of the $k$ 
smallest eigenvalues are entanglement monotones [\Vpure].  The same 
information---in somewhat less familiar, but more algebraically 
convenient form---is contained in the coefficients of the 
characteristic polynomial of the reduced density matrix.  These 
coefficients are polynomials in the components of the state vector and 
their complex conjugates.  They generate the ring of polynomial 
functions invariant under the action of local unitary transformations; 
thus they completely classify two particle pure state entanglement.

As the number of particles $n$ increases, however, the number of 
independent invariants---measures of entanglement---grows 
exponentially.  Complete classification rapidly becomes impractical.
Our goal in this Letter is more modest:  we seek a measure of 
entanglement which is {\sl scalable}, \ie, which is defined for any
number of particles; which is easily calculated; and which provides
physically relevant information.  We concentrate on the case of 
spin-${1\over2}$ particles ({\sl qubits\/}) and begin by defining a 
family (parameterized by $n$) of functions on $\Htwon$.  We show that 
each function is a measure of entanglement, vanishing exactly on 
product states.  Next we evaluate this measure for several example 
states which illustrate its properties, most importantly that it 
measures {\sl global\/} entanglement.  This is perhaps best 
exemplified by its values on eigenstates of the antiferromagnetic 
Hamiltonian, for which we show that it is maximal only on the ground 
state.  In a less traditional context, quantum computation relies 
heavily on multiparticle entangled states, particularly for error 
correction.  We show that quantum error correcting code states also 
maximize our measure of entanglement.  Finally, we illustrate its use 
in a dynamical setting, tracking the evolution of entanglement during 
a specific quantum computation.

The Hilbert space $\Htwon$ of $n$ qubits has a basis labelled by the 
$2^n$ $n$-bit strings:  $|b_1 \ldots b_n\rangle$, $b_j \in \{0,1\}$.  
For $b \in \{0,1\}$, define
$$
\imath_j(b) |b_1 \ldots b_n\rangle 
 = \delta_{bb_j} |b_1 \ldots \widehat{b_j} \ldots b_n\rangle,
$$
where\ \ $\widehat{\phantom{a}}$\ \ denotes absence.  We extend 
$\imath_j$ by linearity to be a map $\C^2 \otimes \Htwon \to \Hless$.
For $u,v \in \Hless$ we can write $u = \sum u_x |x\rangle$ and 
$v = \sum v_y |y\rangle$, where $0 \le x,y < 2^{n-1}$ are 
$(n-1)$-bit strings.  Next, let
$$
D(u,v) = \sum_{x<y} |u_x v_y - u_y v_x|^2,
$$
the norm-squared of the wedge product of $u$ and $v$.  Finally, for 
$\psi \in \Htwon$, define
$$
Q(\psi) = {4\over n}\sum_{j=1}^n 
          D\bigl(\imath_j(0)\psi,\imath_j(1)\psi\bigr).
$$
As we will see shortly, the $4/n$ factor provides a convenient 
normalization for $Q$.

\noindent\Proposition\ 1.  {\sl For each $n \in \Z_{\ge2}$, 
$Q : \Htwon \to \R$ is a measure of entanglement.}

\noindent{\sl Proof}.  For $u,v \in \Hless$, $D(u,v)$ is invariant 
under U$(2^{n-1})$.  A transformation of the $j^{\rm th}$ qubit in 
$\psi \in \Htwon$ multiplies the $j^{\rm th}$ summand in $Q$ by the 
norm squared of its determinant.  Thus each summand is invariant under 
TSL$(2) \times {\rm U}(2^{n-1})$, where T denotes the unit scalars.  
The intersection of these groups for all the summands is U$(2)^n$, 
\ie, the local unitary transformations.                \hfill\endproof

The most basic property that a measure of entanglement can have is to
identify completely unentangled, \ie, product states.  $Q$ has this
property:

\noindent\Proposition\ 2.  {\sl $Q(\psi) = 0$ iff $\psi$ is a product
state.}

\noindent{\sl Proof}.  Two vectors $u,v \in \Hless$ are linearly 
dependent iff $D(u,v) = 0$.  Thus $Q(\psi) = 0$ implies the existence 
of $\alpha_j \in \C$ such that 
$\imath_j(1)\psi = \alpha_j\imath_j(0)\psi$ for all $1 \le j \le n$.
In particular,
$$
\eqalign{
\psi 
 &= |0\rangle \otimes \imath_1(0)\psi + 
    |1\rangle \otimes \imath_1(1)\psi                              \cr
 &= (|0\rangle + \alpha_1|1\rangle) \otimes \imath_1(0)\psi        \cr
 &= (g \otimes I) \cdot (|0\rangle \otimes \psi')                  \cr
}
$$
for some $g \in {\rm SU}(2)$, $\psi' \in \Hless$.  By Proposition~1, 
$Q$ is invariant under the local unitary transformation $g \otimes I$, 
so 
$$
\eqalign{
0 &= Q(\psi) = Q(|0\rangle \otimes \psi')                          \cr
  &= 0 + \sum_{j=2}^n D\bigl(\imath_j(0)[|0\rangle \otimes \psi'],
                             \imath_j(1)[|0\rangle \otimes \psi']
                       \bigr)                                      \cr
  &= Q(\psi').                                                     \cr
}
$$
Then, by induction, $\psi$ is a product state.

Conversely, if $\psi$ is a product state, then for all
$1 \le j \le n$, $\imath_j(0)\psi$ is parallel to $\imath_j(1)\psi$.
Thus $Q(\psi) = 0$.                                    \hfill\endproof

Having demonstrated that $Q$ vanishes on product states, we should now
calculate it for some entangled states.  First consider the EPR-Bohm
[\EPR,\Bohm] state $(|01\rangle - |10\rangle)/\sqrt{2}$, or 
equivalently, $\gamma_2 = (|00\rangle + |11\rangle)/\sqrt{2}$.  It is
straightforward to calculate:
$$
Q(\gamma_2) 
 = 2\cdot{4\over2} \Bigl[\det {1\over\sqrt{2}}\pmatrix{ 1 & 0 \cr
                                                        0 & 1 \cr
                                                      }
                   \Bigr]^2
 = 1.
$$
Next, the three qubit GHZ-Mermin [\GHZ,\Mermin] state is
$\gamma_3 = (|000\rangle + |111\rangle)/\sqrt{2}$.  Calculating the 
invariant, we again find:
$$
Q(\gamma_3) 
 = 3\cdot{4\over3} \Bigl[\det {1\over\sqrt{2}}\pmatrix{ 1 & 0 \cr
                                                        0 & 1 \cr
                                                      }
                   \Bigr]^2
 = 1.
$$
Finally, it is now clear that for the $n$ qubit state
$\gamma_n = (|0\ldots0\rangle + |1\ldots1\rangle)/\sqrt{2}$,
$Q(\gamma_n) = 1$.  These examples demonstrate that the $4/n$ factor 
provides a natural normalization for $Q$.  

\noindent\Proposition\ 3.  {\sl With this normalization, 
$0 \le Q \le 1$.}

\noindent{\sl Proof}.  Since $D(u,v)$ is the norm-squared of the wedge 
product of the two vectors $u$ and $v$, it is bounded above by
$\|u\|^2 \|v\|^2$, which takes its maximal value of ${1\over4}$ when
$\|u\|^2 = \|v\|^2 = {1\over2}$ for vectors 
$\psi = |0\rangle\otimes\imath_1(0)\psi + 
        |1\rangle\otimes\imath_1(1)\psi$ with $\|\psi\|^2 = 1$.  Since
there are $n$ summands in $Q$, it is bounded above by 
$n\cdot{4\over n}\cdot{1\over4} = 1$.                  \hfill\endproof

Proposition~2 and the calculations above show that these bounds are 
saturated on, respectively, product states and the entangled states
$\gamma_n$.  Of course, $Q$ does take other values:  Under the action 
of ${\rm U}(2) \times {\rm U}(2) \times {\rm U}(2)$ the Hilbert space 
for three qubits, $\C^2 \otimes \C^2 \otimes \C^2$, decomposes into 
multiple orbits [\LindenPopescu--\threequbits].  
Representative states are
$|000\rangle$ (product states), 
$|0\rangle \otimes \gamma_2 = 
 |0\rangle(|00\rangle + |11\rangle)/\sqrt{2}$ (and cyclic 
permutations),
$(|100\rangle + |010\rangle + |001\rangle)/\sqrt{3}$, and
$\gamma_3$.  By Proposition~2, $Q(|000\rangle) = 0$.  By Proposition~2
and the calculation above,
$Q(|0\rangle \otimes \gamma_2) = {4\over3}\cdot{1\over2} = {2\over3}$.  
A straightforward calculation gives 
$Q\bigl((|100\rangle + |010\rangle + |001\rangle\bigr)/\sqrt{3}) = 
{8\over9}$.  
And we have already calculated $Q(\gamma_3) = 1$.  Thus for three 
qubits our measure of entanglement behaves in the way we would want, 
decreasing through states we would consider successively less globally 
entangled (and taking different values on each of these states, unlike 
the `tangle' [\CKW], for example, which vanishes on all but 
$\gamma_3$).  In fact, on three qubits $Q$ is an entanglement monotone 
and numerical evidence indicates that this is true in general [\mono].

The traditional context in which globally entangled multiparticle 
states occur is lattice spin systems.  Consider, for example, the one
dimensional spin-${1\over2}$ Heisenberg antiferromagnet 
[\Dirac,\Heisenberg] on a lattice of size $n$, with periodic boundary 
conditions, defined by the Hamiltonian
$$
H_n = \sum_{j=1}^n X_j X_{j+1} + Y_j Y_{j+1} + Z_j Z_{j+1},
$$
where the subscripts are to be interpreted mod $n$, and $X$, $Y$, $Z$
denote the Pauli matrices $\sigma_x$, $\sigma_y$, $\sigma_z$, 
respectively.  $H_n$ commutes with $S_z = \sum Z_j$, so the 
eigenstates of $H_n$ can be labelled by their total spin $S_z$, \ie, 
each eigenstate of $H_n$ is a superposition of basis vectors 
$|b_1\ldots b_n\rangle$ with $|\{j\mid b_j = 1\}| = s$ for some fixed 
$0 \le s \le n$.  When $s = 1$, the translation invariance of $H_n$ 
implies that the eigenstates are plane waves
$$
\psi_n^{(k)} 
 = {1\over\sqrt{n}}\sum_{j=0}^{n-1} e^{ikj}|0\ldots010\ldots0\rangle,
$$
where the $j^{\rm th}$ summand has a single 1 at the $j^{\rm th}$ bit
and the wave number $k = 2\pi m/n$ for some integer $0\le m \le n-1$.
For $n = 3$ these plane waves are equivalent under local unitary 
transformations to the state 
$(|100\rangle + |010\rangle + |001\rangle)/\sqrt{3}$, for which we
calculated $Q = {8\over9}$ above.  In fact, for arbitrary $n$ the 
entanglement of these plane waves is simply
$$
Q\bigl(\psi_n^{(k)}\bigr) = {4\over n}\cdot n\cdot (n-1){1\over n^2}
                          = {4(n-1)\over n^2}.
$$

For $s > 1$ the eigenstates of $H_n$ can be computed using the Bethe
{\it Ansatz\/}; the ground state has $s = n/2$ (for even $n$) 
[\Bethe].  Using the translation invariance of these eigenstates we
can evaluate $Q$ easily.  The result is 
$$
Q(S_z = s {\rm\ eigenstate\ of\ }H_n)
 = {4\over n}\cdot n\cdot{n-1\choose s}{n-1\choose s-1}
   {n\choose s}^{-2}
 = {4s(n-s)\over n^2}.
$$
Notice that for the ground state this entanglement measure is maximal,
$Q = 1$.  This result contrasts with O'Connor and Wootters' 
calculations of the `concurrence' $C$ in these states 
[\OConnorWootters]:  they find that $C$ is not maximal on the ground
state, but rather for $s/n \approx 0.3$ as $n \to \infty$.  This 
difference is due to the fact that $C$ is really a measure of two 
particle entanglement, even when generalized to multiparticle states, 
while $Q$ is a global measure of multiparticle entanglement.

Highly entangled multiparticle states also occur in the relatively new
context of quantum error correcting codes. In fact, the code subspace
for an additive code can be described as the space of ground states of
the Hamiltonian formed by the sum of the stabilizers [\Gottesman].  
For example, the code subspace for a 5 qubit code {\bf [5,1,3]} 
encoding 1 qubit against single bit errors [\BDSW,\LMPZ] is the space 
of ground states of the translation invariant Hamiltonian on a one 
dimensional lattice of five qubits:
$$
H_{[5,1,3]} = \sum_{j=1}^5 X_j Z_{j+1} Z_{j+2} X_{j+3},
$$
where the subscripts are to be interpreted mod 5.  The space of ground
states is two dimensional---which is why it can encode 1 qubit.  A
basis is 
$$
\eqalign{
|0\rangle &\mapsto
 [|00000\rangle - (|11000\rangle + {\rm cyc.})
                + (|10100\rangle + {\rm cyc.})
                - (|11110\rangle + {\rm cyc.})
 ]/4                                                               \cr
|1\rangle &\mapsto
 [|11111\rangle - (|00111\rangle + {\rm cyc.})
                + (|01011\rangle + {\rm cyc.})
                - (|00001\rangle + {\rm cyc.})
 ]/4,                                                              \cr
}
$$
where ``cyc.''\ indicates cyclic permutations.  From these equations 
it is straightforward to calculate that $Q = 1$ for all states in this 
code space.  Here the difference from the concurrence is even more 
dramatic:  $C$ vanishes on the code subspace since tracing over all 
but two qubits leaves a reduced density matrix proportional to the 
identity [\CGL].

Shor's original 9 qubit code protecting 1 qubit against single qubit
errors [\Shor] can also be described as the ground state subspace of a
lattice Hamiltonian---for a lattice triangulating $\R P^2$ [\rpp].  In
this case a basis for the code space is
$$
\eqalign{
|0\rangle &\mapsto (|000\rangle + |111\rangle)^{\otimes 3}/3       \cr
|1\rangle &\mapsto (|000\rangle - |111\rangle)^{\otimes 3}/3.      \cr
}
$$
Calculating $Q$ for the states in this subspace we find that again it
is maximal, despite the fact that these states decompose into products
of three qubit factors.  So $Q$ does not distinguish all sub-global 
entanglements; this is a consequence of using a single invariant.  
Finer resolution requires a more complete set of invariants, and in 
general, higher degree polynomials 
[\LindenPopescu,\GRB,\Sudbery,\threequbits,\BarnumLinden].

\moveright\secondstart\vtop to 0pt{\hsize=\halfwidth
\vskip -1\baselineskip
$$
\epsfxsize=\halfwidth\epsfbox{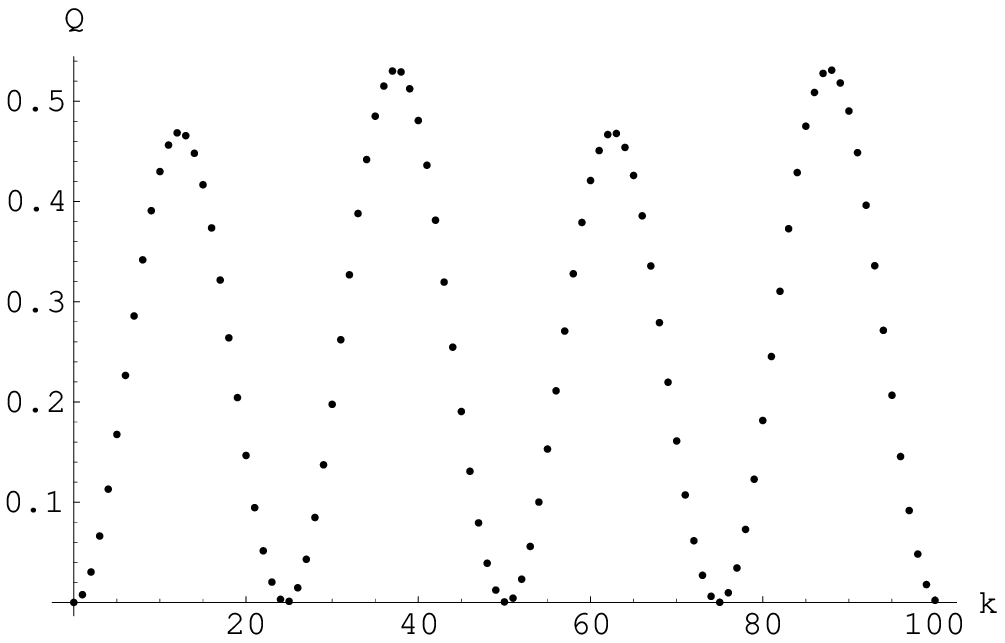}
$$
\vskip -0.75\baselineskip
\eightpoint{%
\noindent{\bf Figure~1}.  Entanglement in Grover's algorithm for 10
qubits as a function of number of iterations.
}}
\vskip -\baselineskip
\parshape=14
0pt \halfwidth
0pt \halfwidth
0pt \halfwidth
0pt \halfwidth
0pt \halfwidth
0pt \halfwidth
0pt \halfwidth
0pt \halfwidth
0pt \halfwidth
0pt \halfwidth
0pt \halfwidth
0pt \halfwidth
0pt \halfwidth
0pt \hsize
Nevertheless, as we have seen, $Q$ provides useful information about
global entanglement in certain contexts.  Furthermore, in dynamical 
problems, $Q$ quantifies the evolution of entanglement.  Consider
Grover's algorithm [\Grover], for example:  Given 
$a_j \in \{0,1\}$, $1 \le j \le n$, define
$$
U_a|b_1\ldots b_n\rangle 
 = (-1)^{\Pi\delta_{a_jb_j}} |b_1\ldots b_n\rangle
$$
and then extend $U_a$ by linearity to a map $\Htwon \to \Htwon$.  The
goal of Grover's algorithm is to convert an initial state of $n$ 
qubits, say $|0\ldots0\rangle$, to a state with probability bounded 
above ${1\over2}$ of being in the state $|a_1\ldots a_n\rangle$, using 
$U_a$ the fewest times possible.  Grover showed that it can be done 
with $O(\sqrt{2^n})$ uses of $U_a$ by preparing the state
$$
{1\over\sqrt{2^n}} \sum_{x=0}^{2^n - 1} |x\rangle 
 = H^{\otimes n} |0\ldots0\rangle
\quad{\rm where}\quad
H = {1\over\sqrt{2}}\pmatrix{1 & 1\cr
                             1 &-1\cr
                            },
$$
and then iterating the transformation 
$H^{\otimes n}U_0H^{\otimes n}U_a$ on this state [\Grover].  The 
initial state is a product state, as is the target state 
$|a_1\ldots a_n\rangle$, but intermediate states $\psi(k)$ are 
entangled for $k > 0$ iterations.  We can evaluate $Q$ on these states
to quantify this entanglement:
$$
Q\bigl(\psi(k)\bigr) 
 = 4\Bigl({N\over2}-1\Bigr){\cos^2\theta_k\over N-1}
    \Bigl(\sin\theta_k - {\cos\theta_k\over\sqrt{N-1}}\Bigr)^2,
$$
where $\theta_k = (2k+1)\csc^{-1}(\sqrt{N})$ and $N = 2^n$.  The 
results are plotted in Figure~1 for $n = 10$:  the entanglement 
oscillates, first returning to close to 0 at
$$
k = \Bigl[\!\!\Bigl[{1\over2}\Bigl({\pi\over 2\csc^{-1}(\sqrt{N})} - 1
                             \Bigr)
    \Bigr]\!\!\Bigr]
  \sim \Bigl[\!\!\Bigl[{\pi\over4}\sqrt{N}\Bigr]\!\!\Bigr]
\quad{\rm as}\quad N\to\infty, 
$$
where $[\![\cdot]\!]$ denotes `closest integer to'; this is when the 
probability of measuring $|a_1\ldots a_n\rangle$ is first close to 1.

Three qubit states, eigenstates of lattice Hamiltonians, quantum error
correcting code subspaces, and the intermediate states in Grover's
algorithm all illustrate how a measure of multiparticle entanglement 
such as $Q$ provides insight into global properties of quantum 
multiparticle systems.  While $Q$ has the satisfactory properties of 
Propositions~2 and 3, is an entanglement monotone on three qubits, and 
is a straightforwardly computable polynomial, it is in no sense a 
unique measure of multiparticle entanglement.  A more complete (but 
still partial) characterization can be obtained by also using some of 
the other measures which have been proposed, like the concurrence 
[\OConnorWootters], the closely related $n$-tangle [\WongChristensen], 
the Schmidt rank [\EisertBriegel], the negativity [\VidalWerner], 
\etc\ \ Each emphasizes a specific feature of multiparticle 
entanglement and describes a different physical property.  We 
anticipate that multiparticle entanglement measures---whose current 
development is largely motivated by quantum computation---will 
contribute to the understanding of the physics of quantum 
multiparticle systems more generally [\Preskill--\ABV].

\medskip
\noindent{\bf Acknowledgements}

\noindent This work was supported in part by the National Security 
Agency (NSA) and Advanced Research and Development Activity (ARDA) 
under Army Research Office (ARO) contract number DAAG55-98-1-0376.

\medskip
\global\setbox1=\hbox{[00]\enspace}
\parindent=\wd1

\noindent{\bf References}
\vskip10pt

\parskip=0pt
\item{[\SchrodingerA]}
\schrodinger,
``{\it Die gegenw\"artige Situation in der Quantenmechanik\/}'',
\Naw\ {\bf 23} (1935) 807--812; 823--828; 844--849.

\item{[\EPR]}
A. Einstein, B. Podolsky and N. Rosen,
``Can quantum-mechanical description of physical reality be 
considered complete?'',
\PR\ {\bf 47} (1935) 777--780.

\item{[\BBPS]}
C. H. Bennett, H. J. Bernstein, S. Popescu and B Schumacher,
``Concetrating partial entanglement by local operations'',
\PRA\ {\bf 53} (1996) 2046--2052.

\item{[\Vmono]}
\vidal,
``Entanglement monotones'',
\JMO\ {\bf 47} (2000) 355--376.

\item{[\Vpure]}
\vidal,
``Entanglement of pure states for a single copy'',
\PRL\ (1998) 1046--149.

\item{[\Bohm]}
D. Bohm,
{\sl Quantum Theory\/}
(New York:  Prentice-Hall 1951).

\item{[\GHZ]}
D. M. Greenberger, M. A. Horne and A. Zeilinger,
``Going beyond Bell's theorem'',
in M. Kafatos, ed.,
{\sl Bell's Theorem, Quantum Theory and Conceptions of the
  Universe\/}
(Boston:  Kluwer 1989) 69--72.

\item{[\Mermin]}
N. D. Mermin,
``What's wrong with these elements of reality'',
\PhT\ {\bf 43} (June 1990) 9--11.

\item{[\LindenPopescu]}
N. Linden and S. Popescu,
``On multi-particle entanglement'',
\FortP\ {\bf 46} (1998) 567--578.

\item{[\GRB]}
M. Grassl, M. R\"otteler and T. Beth,
``Computing local invariants of qubit systems'',
\PRA\ {\bf 58} (1998) 1833--1839.

\item{[\CarteretSudbery]}
H. A. Carteret and A. Sudbery,
``Local symmetry properties of pure 3-qubit states'',
\JPA\ {\bf 33} (2000) 4981--5002.

\item{[\Sudbery]}
A. Sudbery,
``On local invariants of pure three-qubit states'',
\JPA\ {\bf 34} (2001) 643--652.

\item{[\AACJLT]}
A. Ac{\'\i}n, A. Andrianov, L. Costa, E. Jan\'e, J. I. Latorre and 
R. Tarrach,
``Generalized Schmidt decomposition and classification of 
  three-quantum-bit states'',
\PRL\ {\bf 85} (2000) 1560--1563.

\item{[\AAJT]}
A. Ac{\'\i}n, A. Andrianov, E. Jan\'e and R. Tarrach,
``Three-qubit pure-state canonical forms'',
{\tt quant-ph/0009107}.

\item{[\threequbits]}
\dn,
``Invariants for multiple qubits I:  the case of 3 qubits'',
UCSD preprint (2001).

\item{[\CKW]}
V. Coffman, J. Kundu and W. K. Wootters,
``Distributed entanglement'',
\PRA\ {\bf 61} (2000) 052306.

\item{[\mono]}
\dn,
in preparation.

\item{[\Dirac]}
P. A. M. Dirac,
``On the theory of quantum mechanics'',
\PRSLA\ {\bf 112} (1926) 661--677.

\item{[\Heisenberg]}
W. Heisenberg,
``{\it Zur Theorie des Ferromagnetismus\/}'',
\ZP\ {\bf 49} (1928) 619--636.

\item{[\Bethe]}
H. A. Bethe,
``{\it Zur Theorie der Metalle.  I.  Eigenwerte und Eigenfunktionen
       der linearen Atomkette}'',
\ZP\ {\bf 71} (1931) 205--226.

\item{[\OConnorWootters]}
K. M. O'Connor and W. K. Wootters,
``Entangled rings'',
{\tt quant-ph/0009041}.

\item{[\Gottesman]}
\gottesman,
{\sl Stabilizer Codes and Quantum Error Correction},
Caltech Ph.D. thesis, physics (1997),
{\tt quant-ph/9705052}.

\item{[\BDSW]}
C. H. Bennett, D. P. DiVincenzo, J. A. Smolin and W. K. Wootters,
``Mixed-state entanglement and quantum error correction'',
\PRA\ {\bf 54} (1996) 3824--3851.

\item{[\LMPZ]}
R. Laflamme, C. Miquel, J. P. Paz and W. H. Zurek,
``Perfect quantum error correction code''.
\PRL\ {\bf 77} (1996) 198--201.

\item{[\CGL]}
R. Cleve, D. Gottesman and H.-K. Lo,
``How to share a quantum secret'',
\PRL\ {\bf 83} (1999) 648--651.

\item{[\Shor]}
P. W. Shor,
``Scheme for reducing decoherence in quantum computer memory'',
\PRA\ {\bf 52} (1995) R2493--R2496.

\item{[\rpp]}
\md,
``Projective plane and planar quantum codes'',
\FoCM\ {\bf 1} (2001) 325--332.

\item{[\BarnumLinden]}
H. Barnum and N. Linden,
``Monotones and invariants for multi-particle quantum states'',
{\tt quant-ph/0103155}.

\item{[\Grover]}
L. K. Grover,
``A fast quantum mechanical algorithm for database search'',
in 
{\sl Proceedings of the 28th Annual ACM Symposium on the Theory of
Computing}, Philadelphia, PA, 22--24 May 1996
(New York:  ACM 1996) 212--219.

\item{[\WongChristensen]}
A. Wong and N. Christensen,
``A potential multiparticle entanglement measure'',
{\tt quant-ph/0010052}.

\vfill\eject

\item{[\EisertBriegel]}
J. Eisert and H. J. Briegel,
``Quantification of multi-particle entanglement'',
{\tt quant-ph/0007081}.

\item{[\VidalWerner]}
\vidal\ and R. F. Werner,
``A computable measure of entanglement'',
{\tt quant-ph/ 0102117}.

\item{[\Preskill]}
J. Preskill,
``Quantum information and physics:  some future directions'',
\JMO\ {\bf 47} (2000) 127--137.

\item{[\BriegelRaussendorf]}
H. J. Briegel and R. Raussendorf,
``Persistent arrays of interacting particles'',
{\tt quant-ph/0004051}.

\item{[\ABV]}
M. C. Arnesen, S. Bose and V. Vedral,
``Natural thermal and magnetic entanglement in 1D Heisenberg model'',
{\tt quant-ph/0009060}.

\bye